\documentstyle[a4p,12pt,cite,epsfig]{article}
\begin{document}
\begin{titlepage}
-\\
\vspace{3cm}
\begin{center}
{\large\bf The energy dependence of the chaoticity $\lambda$  parameter\\
\vspace{1.5mm}
from BEC of $\pi$-pairs produced in $pp$ collisions}\\
\vspace{10mm}

{\bf  Gideon Alexander$^{a,}$\footnote{Email: gideona@post.tau.ac.il}
and Vitalii A. Okorokov$^{b,}$\footnote{Email: VAOkorokov@mephi.ru; Okorokov@bnl.gov}}\\
\vspace{4mm}
\end{center}
\centering{\it a) School of Physics and Astronomy,
Raymond and Beverly Sackler Faculty of Exact Sciences,
Tel-Aviv University, 69978 Tel-Aviv, Israel\\
b) National Research Nuclear University MEPhI (Moscow Engineering
Physics Institute), 115409 Moscow, Russia} \vspace{10mm}

\begin{abstract}
The $\sqrt{s_{pp}}$ behavior of the chaoticity parameter
$\lambda$, derived from Bose\,--\,Einstein Correlations (BEC) of
pion-pairs produced in $pp$ collisions, is investigated.
Considered are the one and three dimensions (1D, 3D) of the BEC
analyzed in terms of a Gaussian and/or Exponential distributions.
A marked difference is observed between the $\lambda$ dependence
on energy in the 1D and the 3D analyzes. The experimental data are
examined in terms of the relation between the pion cluster of
sources and the BEC dimension R which in turn are deduced from the
charged outgoing particle multiplicity. While in this approach
the general decrease with energy of the 1D $\lambda$ is accounted for
it fails to represent the few 3D
$\lambda$ data which are seen to remain constant with energy
above $\sim$200 GeV.
\end{abstract}
\vspace{5mm}
\begin{center}
\today
\end{center}
\end{titlepage}

\setcounter{footnote}{0}

\section{Introduction}
\label{sec1}

\hspace{6mm}With the recent operation of the Large Hadron Collider (LHC) at
CERN the opportunity to study the Bose\,--\,Einstein Correlations (BEC)
of identical bosons at very high hadron-hadron collision energy
has been opened \cite{cms1,cms2,cms3,alice1,alice2,alice4}. In
particular the energy dependence of the BEC dimension $R$ has
recently been investigated and was found to increase with
log$(\sqrt{s_{NN}})$ both in proton-proton ($pp$) and in heavy ion
($AA$) collisions \cite{alexander,gaitai,okorokov}. In the present
work we investigate the energy behavior of the chaoticity
parameter $\lambda$ in $pp$ collisions which determines the
strength of the measured BEC effect. To this end we utilize the
relation between the BEC dimension $R$, the number of pion source clusters
and $\lambda$. We further stipulate that the pion
source clusters are proportional to the average charged particle
multiplicity produced in the hadron reactions.\\

The BEC is measured in terms of the two identical particle
correlation function
\begin{equation}
C(p_1,p_2)\ =\ \frac{\rho(p_1,p_2)}{\rho_0(p_1,p_2)}\ ,
\label{bec}
\end{equation}
where $p_1$ and $p_2$ are the 4-momenta of the two hadrons, $\rho$
is the two particle density function and $\rho_0$ is the two
particle density function in the absence of the BEC effect. This
$\rho_0$ is often referred to as the reference sample against
which the BEC is measured. There are several ways to construct
$\rho_0$ which were adopted by the different BEC experiments
\cite{review}. These and the different background
conditions and the variety of BEC analysis methods should be appraised
when their physics implications are determined.
Throughout this work we assume that the BEC
background is well accounted for and that one can safely ignore
the influence of the long range correlations on the $\lambda$
properties.

\section{The one dimension BEC analysis}
\hspace{6mm}Among the various BEC representations one of the frequently used
in the one dimension (1D) analysis of hadrons emerging from a
sphere volume, is the Goldhaber parametrization of a static
Gaussian source in the plane-wave approach \cite{goldhaber},
namely
\begin{equation}
C_{Gauss}(Q)\ =\ 1+\lambda_{Gauss}e^{-Q^2R^2_{Gauss}}\ ,
\label{gauss}
\end{equation}
which assumes for the particles emitter a
spherical volume with a radial Gaussian
distribution. The second often used
parametrization, which assumes a radial Lorentzian distribution of
the source, is given by
\begin{equation}
C_{Expo}(Q)\ =\ 1+\lambda_{Expo}e^{-QR_{Expo}}\ ,
\label{expo}
\end {equation}
which generally was found at low $Q$ values, e.g.
$Q \leq 0.1$ GeV, to fit beter the measured
BEC distribution than the
the Gaussian parametrization \cite{atlasbec}.
In both representations ${Q^2=-(p_1-p_2)^2}$ is the difference
squared of the 4-momentum vectors of the two correlated identical
bosons. The $\lambda$ factor, also known as the chaoticity
parameter, lies in the range  between 0 and 1.\\

\begin{figure}[ht]
\centering{\psfig{file=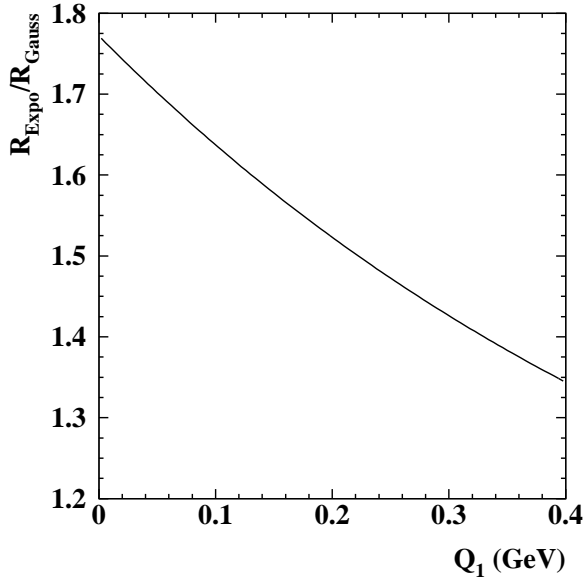,height=8.0cm,
bbllx=46pt,bblly=150pt,bburx=552pt,bbury=682pt}} \caption{\small
$R_{Expo}/R_{Gauss}$ as a function of the lower integration limit
$Q_1$ in Eq. (\ref{rcrg1}) and the upper integration value is
$\geq$ 2 GeV.}
\label{qmin}
\end{figure}

In the 1D analysis the relation between $R_{Gauss}$ and $R_{Expo}$
dimensions can be evaluated from the requirement that the first
$Q$ moment in a given BEC distribution will be equal whether it is
treated by a fit to a Gaussian distribution or to an Exponential one,
namely
\begin{equation}
\frac{\displaystyle
\int_{Q_1}^{\infty}{Qe^{-R^2_{Gauss}Q^2}dQ}}{\displaystyle
\int_{Q_1}^{\infty}{e^{-R^2_{Gauss}Q^2} dQ}}= \frac{\displaystyle
\int_{Q_1}^{\infty}{Qe^{-R_{Expo}Q}dQ}}{\displaystyle
\int_{Q_1}^{\infty}{e^{-R_{Expo}Q} dQ}}\ .
\label{rcrg1}
\end{equation}
This relation remains essentially the same as long as the upper
integration value is higher than 2 GeV. The dependence of
$R_{Expo}/R_{Gauss}$ on the lower integration limit
$Q_1$ is shown in Fig. \ref{qmin}. In the case that
$Q_1=0$ GeV one obtains the known relation
\begin{equation}
R_{Gauss}=\frac{R_{Expo}}{\sqrt{\pi}}\ .
\label{rr}
\end{equation}

A relation between $\lambda_{Gauss}$ and $\lambda_{Expo}$
can in principle be estimated by considering the relations
\begin{equation}
\int^{Q_2}_{Q_1}\lambda_{Expo}e^{-QR_{Expo}}dQ\simeq
\int^{Q_2}_{Q_1}\lambda_{Expo}e^{-QR_{Gauss}\sqrt{\pi}}dQ\simeq
\int^{Q_2}_{Q_1}\lambda_{Gauss}e^{-Q^2R_{Gauss}^2}dQ\ ,
\label{good}
\end{equation}
where in the range between $Q_1$ and $Q_2$ GeV the Gaussian and
the Exponential BEC parametrizations fit equally well the measured
$Q$ distribution. In this case for the values of $Q_1=0$ and $Q_2=\infty$ GeV
one finds that
\begin{equation}
\lambda_{Gauss}\ =\ \frac{2\lambda_{Expo}}{\pi}\ .
\label{lamratio}
\end{equation}

Experimentally it has been found that the chaoticity
parameter depends on several of the data properties such as
the average particle pair transverse momentum, $\langle
k_{T}\rangle$, and in particular on the outgoing
charged particle multiplicity $N_{ch}$.
Hence the additional division of the experimental measured
$\lambda$ data into different categories is advisable
before attempting a meaningful comparison with a model prediction.
For this reason we have here, unlike in a former preliminary study
\cite{ICPPA-2015}, considered two distinct $pp$ collision data
sample, the first without any multiplicity restrictions labeled by
MB (Minimum-Bias events) and the second the High Multiplicity
events, labeled by HM.
The measured values of
$\lambda_{Gauss}^{MB}$ and $\lambda_{Gauss}^{HM}$ are given in
Table \ref{table1}.
\vspace{2mm}

The measured ratios $\lambda_{Expo}/\lambda_{Gauss}$, evaluated from Table \ref{table1},
are significantly different from that of $\pi/2$ given by Eq.
(\ref{lamratio}) as they are lying in the neighborhood of
the value 2. The relation between the parameter pairs
$(R_{Gauss}, \lambda_{Gauss})$ and $(R_{Expo}, \lambda_{Expo})$
obtained in BEC fits to the measured $Q$
distributions are studied in some details in Ref. \cite{note}
where the $\lambda_{Expo}/\lambda_{Gauss}$ obtained ratios are
matched much better to the measured ones inferred from Table \ref{table1}.

\section{The $R$ and $\lambda$ dependence on the $pp$ energy}
\hspace{6mm}It has been shown that in the 1D BEC there exists a
relation between the dimensions $R_{n=1}$ and $R_{n>1}$, where $n$
is the number of independent similar emitting pion source clusters
\cite{rlamn}, namely
\begin{equation}
 R_n = n\frac{\lambda_n}{\lambda_1}R_{n=1}\ .
\label{rs}
\end{equation}
which relates the one sources dimension $R_1$ at low
energies with its value $R_n$ for $n$ identical cluster of sources.
The $\lambda_1$ and $\lambda_n$ parameters are the chaoticity
values respectively for $n=1$ and $n$ clusters. Since we limit
ourselves to the study of
the behavior of the chaoticity parameter with energy
it is sufficient to evaluate the
energy dependence of the ratio $\lambda=\lambda_n/\lambda_1$ .
\vspace{2mm}

Recently the one dimension $R$ dependence on the $pp$ collision
center of mass energy energy was observed to increase
from low energies of less than 100 GeV to those reached by the
LHC at 7 TeV. This rise  was fitted
\cite{alexander} to yield
\begin{equation}
R(s)\ =\ \left\{(1.64\pm 0.11)+(0.14\pm
0.02)\ln(\sqrt{s/s_0})\right\}R_{n=1}\ ,
\label{r1d}
\end{equation}
where $s$ is in TeV$^2$ units and $s_0=1$ TeV$^2$.
\vspace{2mm}
\renewcommand{\arraystretch}{1.2}
\begin{table}[h]
\caption{\small Measured $\lambda$
values obtained from the 1D Bose$-$Einstein correlations of two
identical pion pairs produced in $pp$ collisions using the
Gaussian and/or the Exponential parametrizations according to Eqs.
(\ref{gauss}) and (\ref{expo}). The superscript MB and HM
refer respectively to results from all events and from only high
charged multiplicity events. Whenever the statistical and
systematic errors were reported separately they were added in
quadrature in the table.}
\begin{center}
\begin{tabular}{||cl|c||c|c||c||}
\hline\hline \multicolumn{3}{||c||}{1D BEC analyzes}
& \multicolumn{3}{c||}{The chaoticity $\lambda$ parameter}\cr
\cline{1-6} &Reference & $\sqrt{s}$ (GeV) &
$\lambda_{Gauss}^{MB}$ & $\lambda_{Gauss}^{HM}$ & $\lambda_{Expo}$
\cr \hline \hline &\cite{uribe}& 7.21 & 0.466 $\pm$ 0.015\ & 0.532
$\pm$ 0.013\ &
--- \cr
  &\cite{bailly}&26.0&
0.32 $\pm$ 0.08\ & 0.43 $\pm$ 0.13\ & --- \cr
 &\cite{break}& 31.0  &
0.41 $\pm$ 0.02\ & 0.35 $\pm$ 0.04\ & --- \cr
 &\cite{break}&44.0& 0.40 $\pm$ 0.02\ & 0.42 $\pm$ 0.04\ & --- \cr
 &\cite{akesson-1}&58.1& 0.34 $\pm$ 0.04\
& --- & --- \cr
 &\cite{break}&62.0& 0.43 $\pm$ 0.02\ & 0.42 $\pm$ 0.08\ & --- \cr
 &\cite{akesson-2}&63.0 & 0.39 $\pm$ 0.07\ & --- & 0.77$\pm$
0.07 \cr
 &\cite{breakstone}&63.0&0.45 $\pm$ 0.03\ & --- & --- \cr
 &\cite{agg}&200$^{*}$& 0.35 $\pm$ 0.04\
& 0.36 $\pm$ 0.04\ & --- \cr
 &\cite{alice1}&900& 0.35 $\pm$ 0.03\
& 0.31 $\pm$ 0.03\ & 0.55 $\pm$ 0.05 \cr
 &\cite{atlasbec,sykora}&900& 0.34 $\pm$ 0.03\ & --- & 0.74 $\pm$ 0.11 \cr
 &\cite{cms1,cms2}&900& 0.315 $\pm$
0.014\ & --- & 0.63 $\pm$ 0.03 \cr
 &\cite{cms1}&2360& 0.32 $\pm$ 0.01\
 & --- & 0.66 $\pm$ 0.09 \cr
&\cite{alice3}&7000& 0.65 $\pm$ 0.05\ & 0.66 $\pm$ 0.07\ & --- \cr
  &\cite{atlasbec,sykora}&7000
& 0.33 $\pm$ 0.02\ & 0.25 $\pm$ 0.02\ & 0.53 $\pm$ 0.05 \cr
&\cite{cms2}&7000 &
---
 & --- & 0.62 $\pm$ 0.04\cr
\hline\hline
\multicolumn{6}{l}{$^{*}$\rule{0pt}{10pt}\footnotesize The
relative systematic uncertainty for $\lambda$ is taken to be
equal to the corresponding error associated}\vspace{-2mm} \cr
\multicolumn{6}{l}{\footnotesize with the  BEC radius.} \cr
\end{tabular}
\end{center}
\vspace{-3mm} \label{table1}
\end{table}
\renewcommand{\arraystretch}{1.0}

It has further been shown that the experimental BEC results are
depending only slightly, if at all, on the
rapidity\footnote{Throughout this work we refer to pseudorapidity
by rapidity.}
extent used in the accumulation of
pion-pairs data sample \cite{rapidity}. Thus $R$
is essentially
independent of the rapidity domain used in the experimental BEC
analyzes and as such should also be valid for the results obtained
from particle tracks at the mid rapidity region.
\vspace{2mm}

The energy behavior of $\lambda$ in terms of  Eqs. (\ref{rs})
and (\ref{r1d}) thus requires a solution of the equation
\begin{equation}
\left\{(1.64\pm 0.11)+(0.14\pm
0.02)\ln(\sqrt{s/s_0})\right\}R_{n=1}\ =\ n(s)\lambda R_{n=1}\ ,
\label{extraction}
\end{equation}
where $n(s)$ is the number of source clusters which depends
on energy in the chosen rapidity domain. As has
been found that $R$ increases with the average charged
multiplicity $\langle N_{ch}\rangle$ of the colliding hadrons
and since our aim is to estimate the $\lambda$ dependence on
energy but not on its absolute value, it is sufficient
to require that the number of source clusters is proportional to the
average charged multiplicity.

\subsection{The $\lambda_{1D}$ energy dependence}
\begin{figure}[ht]
\centering{\psfig{file=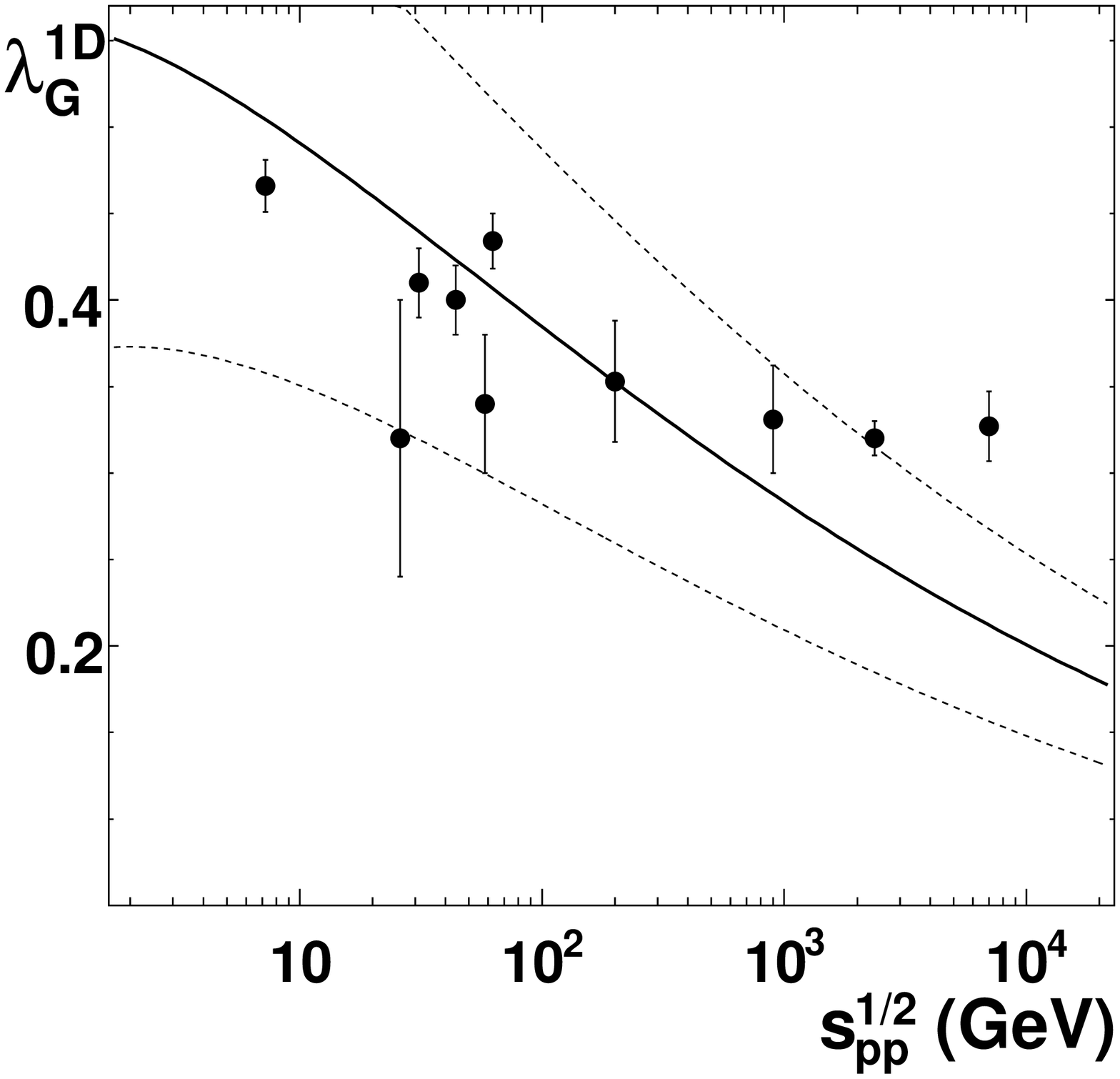,height=8.0cm,
bbllx=0pt,bblly=10pt,bburx=542pt,bbury=552pt}\
{\psfig{file=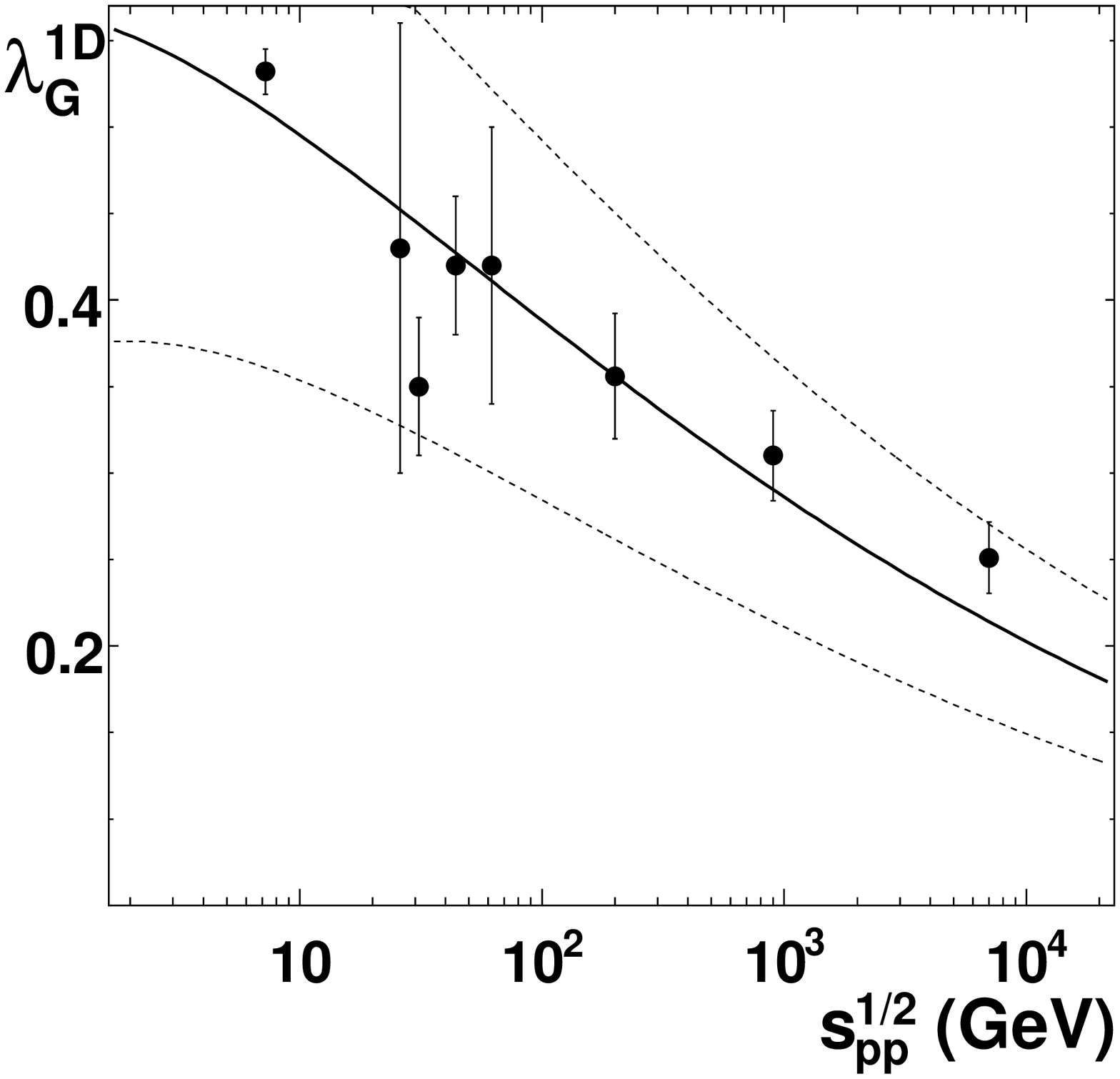,height=8.0cm,
bbllx=0pt,bblly=10pt,bburx=542pt,bbury=552pt}}}
\caption{\small
The 1D two-pion BEC results for  $\lambda_{Gauss}$ as a function
of $\sqrt{s}$.
The expected $\lambda_{Gauss}$ dependence on energy calculated in this work
is shown by the continuous lines normalized to the measured
$\lambda_{Gauss}$ at 200 GeV. The dotted lines represent its
$\pm$1 s.d. limits. Left: Events without a charged multiplicity cut.
The multiple
$\lambda_{Gauss}$ values at 900 was averaged
as well as an average was taken of the $\lambda_{Gauss}$ values
at $\sqrt{s}=62 - 63$ GeV.
Right: Measured $\lambda_{Gauss}$ in high charged
multiplicity events.}
\label{data1d}
\end{figure}

\hspace{6mm}A compilation of $\lambda_{Gauss}$ and $\lambda_{Expo}$ deduced
from the 1D BEC of pion-pairs produced in $pp$ collisions is given
in Table \ref{table1} ordered according to their $pp$ energy. The
measured 1D $\lambda_{Gauss}$ are also plotted in Fig.
\ref{data1d} where the chaoticity values in the energy region of 20
to 60 GeV are seen to be scattered somewhat, most probably due to the
different adopted experimental procedures as pointed out in
Section \ref{sec1}. In spite of this, a general decrease with
energy of the $\lambda_{Gauss}$ values is apparent. Here it should
be noted that the ALICE results at $\sqrt{s}=7$ TeV \cite{alice3},
which are outside the boundary of the figures are quite different
from those of the ATLAS experiment \cite{atlasbec} and also are
far from being part of the general pattern of the $\lambda(s)$
energy dependence.
\vspace{2mm}

To evaluate the $\lambda_{Gauss}$ dependence on energy we follow
the formalism outlined in Ref. \cite{nardi} where the
hadron-hadron collisions is contributed by two components. The
first is the ``hard'' component, with a contribution fraction $x$,
which is due to the number of binary collisions $N_{coll}$, and
the remaining $1-x$ fraction originates from the number of
participants $N_{part}$ referred to as the ``soft '' processes. In
the case of $pp$ collisions one has  $N_{part}=2$ and
$N_{coll}=1$, so that the number of outgoing charged particles per
rapidity unit in $pp$ can be noted as
$\left.(dN_{ch}/{d\eta})\right|_{\eta=0}= n_{pp}$ in accordance
with \cite{nardi}. \vspace{2mm}

In general in hadron collisions the  BEC is analyzed in a rapidity
range symmetric to its central value of $\eta=0$ and it is only
slightly dependent, if at all, on the extent of the rapidity
domain used in the analysis (see, e.g. Refs. \cite{abe,alicerap}).
Thus one should expect  that the charge particle multiplicity
utilized in a  BEC analysis, is approximately proportional to
$\left.(dN_{ch}/{d\eta})\right|_{\eta=0}$. \vspace{2mm}

For the energy dependence of the charged multiplicity
mid-rapidity density we have
considered three log and power series expressions
\cite{abe,EPJPlus-128-45-2013,alice}
given by
\begin{equation}\label{eq:11}
\left.\biggl(\frac{\textstyle dN_{ch}}{\textstyle d\eta}\biggr)\right|_{\eta=0}=\left\{
\begin{array}{lr}
\vspace*{0.35cm} \sum_{i=0}^2a_i\ln^{i}(s/s_{0}),&
\hspace*{0.0cm}{\rm{(a)}}\\ \vspace*{0.35cm}
\sum_{i=0}^2a_i\ln^{i}(s_{a}/s_{0}),&
\hspace*{0.0cm}{\rm{(b)}}\\
a_{0}(s/s_{0})^{a_{1}}.& {\rm{(c)}}
\end{array}
\right.
\label{eta}
\end{equation}
where $s_{a} \equiv (\sqrt{s}-2m_{p})^{2}$ and $m_{p}$ is the proton
mass. The $a_i$ are the free parameters which were determined from the
data to yield the values given in Table \ref{table1.add}. To note is that in Eq.
(\ref{eq:11}) and in the subsequent formulas  $s$ is given in
units of GeV$^2$ and $s_0=1$ GeV$^{2}$, unless otherwise
specifically indicated.
\vspace{2mm}
\renewcommand{\arraystretch}{1.2}
\begin{table}[h]
\caption{\small The free parameter values
obtained for the
rapidity density of charged multiplicity in $pp$ collisions.}
\begin{center}
\begin{tabular}{||l|c||c|c|c||}
\hline\hline Reference & Eq. & $a_{0}$ & $a_{1}$ &$a_{2}$ \cr
\hline \hline \cite{abe} & (11a) & 2.5 $\pm$ 1.0\ & -0.25 $\pm$
0.19\ & 0.023 $\pm$ 0.008\cr \cite{EPJPlus-128-45-2013} & (11b) &
0.39 \ & 0.09 & 0.011 \cr \cite{alice} & (11c) & 0.75 $\pm$ 0.06 \
& 0.114 $\pm$ 0.003& -- \cr \hline\hline
\end{tabular}
\end{center}
\vspace{-3mm} \label{table1.add}
\end{table}
\renewcommand{\arraystretch}{1.0}

\begin{figure}[ht]
\centering{\psfig{file=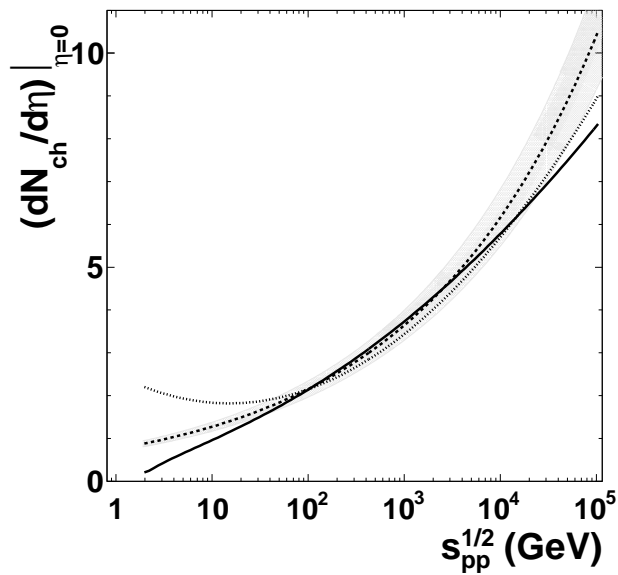,height=8.0cm,
bbllx=0pt,bblly=10pt,bburx=542pt,bbury=552pt}}
\caption{\small
Energy dependence of pseudorapidity density of secondary charged
particles produced in proton-proton and anti-proton-proton
collisions at mid-rapidity ($\eta=0$). The dashed curve
corresponds to the analytic form (\ref{eq:11}a) with parameters
from \cite{abe}, the solid line represents the modified log-series parametrization
(\ref{eq:11}b) from \cite{EPJPlus-128-45-2013} and the dotted curve is
the power function (\ref{eq:11}c) from \cite{alice} with its uncertainty  band.
\label{dNch-pp}}.
\end{figure}

In Fig. \ref{dNch-pp} are shown the energy dependence of the
mid-rapidity charge particle densities according to Eq. (\ref{eq:11})
using their parameter values given in  Table \ref{table1.add}.
As can be seen, the three Eq. (\ref{eq:11}) expressions
agree among themselves in
the energy range from $\sqrt{s} \sim 30$ GeV up to
of 8 TeV and as such do follow well the measured charge multiplicity
density $(dN_{ch}/d\eta)|_{\eta=0}$ in the range
$\sqrt{s}\sim 200$ GeV to 8 TeV. For our analysis we have chosen
the parametrization given by  Eq. (\ref{eq:11}c)
which agrees well with the measured data
further up to 13 TeV and quotes the smallest relative
errors for its components.
Thus we have
\begin{equation}
\left.(dN_{ch}/{d\eta})\right|_{\eta=0}(s)=(0.75 \pm
0.06)(s/s_0)^{0.114 \pm 0.003}\ .
\end{equation}
Inserting $n(s)=\left.(dN_{ch}/{d\eta})\right|_{\eta=0}(s)$ into
Eq. (\ref{extraction}) one obtains the following relation:
\begin{equation}
C\left\{(0.67\pm 0.18)+(0.14 \pm
0.02)\ln(\sqrt{s/s_0})\right\}\ =\ \lambda\left\{(0.75 \pm 0.06)(s/s_{0})^{0.114 \pm 0.003}\right\},
\label{extraction.1}
\end{equation}
where $C$ is a normalization factor.
Solving $\lambda$ from
Eq. (\ref{extraction.1}) one obtains
\begin{equation}
\lambda_{Gauss}(s) \simeq C\frac{(0.89\pm 0.25)+(0.19 \pm
0.03)\ln(\sqrt{s/s_0})}{(s/s_0)^{0.114 \pm 0.003}}.
\label{lamvsen}
\end{equation}
In the present work the normalization of Eq. (\ref{extraction.1})
was determined by requiring that our calculated $\lambda$ will be
equal to the experimentally well measured $\lambda_{Gauss}$ values
at 200 GeV \cite{agg} given in Table \ref{table1}. This yielded
for the MB and  HM data samples respectively nearly the equal values of
$C_{MB}=0.63 \pm 0.11$ and $C_{HM}=0.63 \pm 0.13$.
\vspace{2mm}

The experimentally determined $\lambda$ values
are shown in Fig. \ref{data1d}
as a function of $\sqrt{s}$ for $pp$ collision data
free of charged particle multiplicity limitation (left)
and for only high multiplicity
events (right). The data are compared in both
figures
with our normalized calculated estimations
accompanied by a
$\pm$1 s.d. band limits drawn by the dotted curves.
As can be seen, our calculated  $\lambda_{Gauss}$ behavior
with the $pp$ energy is within 1 s.d in good agreement
with the general decrease with energy of the measured chaoticity
values obtained
from the HM data sample. For the BEC deduced $\lambda_{gauss}$
from the MB data sample
our approach  seem somewhat to deviate  from the data
at $pp$ energies above $\sim$1 TeV.
From this
follows that in $pp$ collisions at $\sqrt{s}=13$
TeV and in the current highest planned LHC energy of $\sqrt{s}=14$
TeV, the expected 1D $\lambda_{Gauss}$ values for HM events should approach
$\sim 0.20$.

\subsection{The $\lambda_{3D}$ energy dependence}
\hspace{6mm}The BEC analysis in three dimensions (3D) is
frequently represented in its Gaussian form by
\begin{equation}
C_{3D}(Q_{long},Q_{out},Q_{side})\ =
\ 1\ +\ \lambda_{3D}
e^{-(R_{long}^2Q_{long}^2+R_{out}^2Q_{out}^2+R_{side}^2Q_{side}^2)}\ ,
\label{threed}
\end{equation}
where the directions $long$, $out$ and $side$ are defined in the
Longitudinal Center of Mass  System (LCMS). (see e.g. Ref.
\cite{review}). The $\lambda_{3D}$ measured values deduced from
BEC carried out in $pp$ collisions
at center of mass energies of 200, 900 and 7000 GeV
are listed in Table \ref{table2}.
At the LHC energies as well as at  $\sqrt{s}=200$ GeV
there are only a qualitative
indication for a smooth decrease of $\lambda_{3D}$ with multiplicity
together with a relative small dependence on $k_{T}$ \cite{alice2}.
\vspace{2mm}

This $k_{T}$-dependence of the chaoticity parameter
is studied in 3D BEC analyzes but not in the Gaussian representation of the
1D BEC studies \cite{agg}.
As only $\lambda_{1D}$ and $\lambda_{3D}$ values derived
at similar, or approximately, experimental conditions can be used
for a meaningful comparison,
the $\lambda_{3D}$ values
given in Table \ref{table2} were those obtained at the lowest  $k_{T}$.
Furthermore for a comparison purpose the chaoticity data shown
in  Table \ref{table2} are those deduced from
the BEC analyzes of the high multiplicity $pp$ collision data.
As can be seen, the $\lambda_{3D}$ values are
higher at the LHC than the corresponding 1D $\lambda_{Gauss}$
obtained at the same $pp$ center of mass energy\footnote{It should
be noted that this conclusion is also valid if the estimation
$\lambda_{3D} \sim 0.49$ 
is used for the minimum bias events at the
LHC energies based on the qualitative information
from \cite{alice2} together with corresponding values for $\lambda_{1D}$
given in Table \ref{table1}.} and the ratio
${\lambda_{3D}}/{\lambda_{1D}}$ indicates
some growth with energy within their uncertainties.
\vspace{2mm}

\renewcommand{\arraystretch}{1.2}
\begin{table}[h]
\caption{\small A comparison between the $\lambda_{3D}$ and
$\lambda_{1D}$ measured in the high multiplicity $pp$ collision events.}
\begin{center}
\begin{tabular}{||cl|c||c|c|c||}
\hline\hline \multicolumn{3}{||c||}{3D BEC in $pp$
collisions}  & \multicolumn{3}{c||}{The measured
$\lambda_{Gauss}$ in HM events}\cr
\cline{1-6} &Reference & $\sqrt{s}$
(GeV) & $\lambda_{1D}$ & $\lambda_{3D}$
&${\lambda_{3D}}/{\lambda_{1D}}$ \cr \hline \hline
 &\cite{agg}&200$^{*}$&  0.35 $\pm$ 0.04\
& 0.42 $\pm$ 0.04\ & 1.20 $\pm$ 0.18  \cr
 &\cite{alice1,alice2}&900& 0.31 $\pm$ 0.03 \
& 0.42 $\pm$ 0.04 & 1.35 $\pm$ 0.18 \cr
&\cite{sykora,alice2}&7000& 0.25 $\pm$ 0.02 \ & 0.42 $\pm$ 0.04&
1.68 $\pm$ 0.21  \cr \hline\hline
\multicolumn{6}{l}{$^{*}$\rule{0pt}{10pt}\footnotesize The
relative $\lambda$ systematic uncertainty is taken to be equal to
the corresponding error of the}\vspace{-2mm} \cr
\multicolumn{6}{l}{\footnotesize BEC dimension.} \cr
\end{tabular}
\end{center}
\vspace{-3mm}
\label{table2}
\end{table}
\renewcommand{\arraystretch}{1.0}

\section{Summary}
\hspace{6mm}The 1D BEC measured values of $\lambda_{Gauss}$ and with them also
the values of  $\lambda_{Expo}$, show a general decrease with
the $pp$ collision energy in particular in the high charged
multiplicity events which does point to
an increase in the coherent pion production.
\vspace{2mm}

The approach adopted here, in which the dimension
$R$ and the multiplicity increase with the $pp$ collision energy
are correlated to the number of source clusters, yield a
decrease of the 1D $\lambda_{Gauss}$ with energy.
The results of this approach agree well
with experimentally 1D BEC deduced $\lambda_{Gauss}$ values obtained
from the high charged
particle multiplicity data up to the $pp$ multi-TeV energy region.
As for the $\lambda_{Gauss}$ obtained
from event samples without any cut on the
outgoing charged particle multiplicity, there is some
discrepancy in the multi-TeV energy region
between the data
and the calculated model expectation
\vspace{2mm}

The chaoticity values extracted from the 3D Bose\,--\,Einstein
correlations are significantly higher than those
obtained in the 1D analyzes, and they seem to remain
essentially constant at the high energy end of the
currently available data.

\subsection*{Acknowledgments}
We would like to thank T. Cs{\"{o}}rg{\'{o}}, C. Pajares and  E.K.G. Sarkisyan for helpful suggestions and comments.

\end{document}